\documentclass{appolb}
\usepackage{epsf}

\begin{document}

\title{Empirics versus RMT in financial cross-correlations}

\author{S. Dro\.zd\.z$^{1,2}$, J. Kwapie\'n$^{1}$, P. O\'swi\c 
ecimka$^{1}$
\address{$^1$Institute of Nuclear Physics, Polish Academy of Sciences,
PL--31-342 Krak\'ow, Poland \\
$^2$Institute of Physics, University of Rzesz\'ow, PL--35-310 Rzesz\'ow,
Poland}}

\maketitle

\begin{abstract}

In order to pursue the issue of the relation between the financial 
cross-correlations and the conventional Random Matrix Theory we analyse 
several characteristics of the stock market correlation matrices like the 
distribution of eigenvalues, the cross-correlations among signs of the 
returns, the volatility cross-correlations, and the multifractal 
characteristics of the principal values. The results indicate that 
the stock market dynamics is not simply decomposable into 'market', 
'sectors', and the Wishart random bulk. This clearly is seen when the time 
series used to construct the correlation matrices are sufficiently long 
and thus the measurement noise suppressed. Instead, a hierarchically 
convoluted and highly nonlinear organization of the market emerges and 
indicates that the relevant information about the whole market is encoded 
already in its constituents.

\end{abstract}


\section{Introduction}

The financial markets represent probably the most complex
structure that is associated with the contemporary civilization.
They involve extremely many constituents, many different space and
time scales and an uncountable number of convoluted factors that
drive the financial dynamics towards a real complexity. Its most
relevant feature is a permanent competitive coexistence of
collectivity and noise. The related quantitative characteristics
can be studied using multivariate ensembles of parameters that
represent dynamics of various financial assets. Due to this
multi-dimensionality the most natural and efficient formal frame
to quantify the whole variety of effects connected with complexity
is in terms of matrices~\cite{drozdz02}. Since the dynamics of
complexity is inherently embedded in noise, the Random Matrix
Theory (RMT)~\cite{wigner51,mehta91} offers an appropriate
reference. Deviations from RMT help to detect real information and
to potentially extract it from what is universal in the RMT sense
and thus practically not very informative. An extremely useful
matrix approach to the financial dynamics is based on using the
correlation matrices formed from (i) the time series representing
the price changes of a certain basket of different assets over the
same period of time or (ii) from the time series representing
different disconnected periods of time (days or weeks) for either
a single asset or an index. The simplest commonly used variant of
RMT to serve as a null hypothesis in these cases corresponds to
the ensemble of Wishart matrices~\cite{wishart28}. The resulting
eigenvalue distribution $\rho(\lambda)$ is then described by the
Marchenko-Pastur formula~\cite{marcenko67,sengupta99} which
confines this distribution within the bounds
\begin{equation}
\lambda_{\rm min}^{\rm max} = 1 + 1 / Q \pm 2 / \sqrt Q.
\label{wishart}
\end{equation}
Here $Q=T/N$ where $N$ is the number of time series of length $T$.
Relating eigenspectra of the empirical $N \otimes N$ stock market
correlation matrices to this formula shows that typically only a
few of the eigenvalues, representing a global or some more local
collective moves within the market, are located sizeably above
$\lambda_{\rm max}$ while the bulk of the empirical eigenvalue
distribution satisfactorily falls within the lower and the upper
bound. This is interpreted as an indication that eigenvectors
associated with the bulk are undistinguishable from noise and thus
carry no information. Such a situation is quite convincing in the
case denoted above as (ii)~\cite{kwapien02,rak06}. The results of
the original study of cross-correlations among the stock market
companies were interpreted analogously~\cite{laloux99,plerou99a}.
A more systematic analysis of this kind of correlations (case (i)
above) shows however~\cite{kwapien06} that they are much more
subtle, the overlap of the bulk with the bounds prescribed by RMT
dissolves as $T$ increases and eigenvectors even from the middle
of the spectrum carry significant information. Below we
recapitulate the most relevant results and provide some further
arguments in favor of the statement that there is nontrivial
information encoded also in the bulk of the eigenvalue spectrum of
the stock market correlation matrix (see also~\cite{malevergne04}). These 
results should be taken care of also in the context of the Markowitz 
optimal portfolio theory~\cite{markowitz52} and for denoising of the 
empirical correlation matrices~\cite{pafka04,burda04}.

\section{Notation}

In the financial context one considers a portfolio $P$ consisting
of a number of securities $X_s, s=1,...,N$ associated with weights
$w_s$ that reflect the fraction of the total capital invested in a
particular security. On the time scale $\Delta t$ the return of
such a portfolio at the $t_j$ instant of time is the weighted sum
\begin{equation}
G_P(j,\Delta t) = \sum_{s=1}^N w_s g_s(j,\Delta t)
\label{portreturn}
\end{equation}
of logarithmic price increments
\begin{equation}
g_s(j,\Delta t)=\ln p_s(t_j+\Delta t) - \ln p_s(t_j) \ \ s=1,...N;
\ j=1,...,T
\label{return}
\end{equation}
of individual securities $X_s$. Each such return can be considered
a product
\begin{equation}
g_s(j) = {\rm sign}_s(j) \times v_s(j)
\label{svreturn}
\end{equation}
of its sign and of its magnitude $v_s(j)$ which measures the volatility.

These time series can be used to create an $N \times T$ data
matrix ${\bf M}$ and then a correlation matrix ${\bf C}$ according
to the formula
\begin{equation}
{\bf C} = (1 / T) {\bf M} {\bf M}^{\rm T}.
\end{equation}
Each element of ${\bf C}$ is thus the Pearson correlation
coefficient $C_{mn}$ between a pair of signals $m$ and $n$. By
solving the eigenvalue problem
\begin{equation}
{\bf C} {\bf x}_i = \lambda_i {\bf x}_i, \ \ i = 1,...,N,
\label{diag}
\end{equation}
this matrix can be transformed to the diagonal form. From the
point of view of investment theories, each eigenvector ${\bf x}_i$
can be considered as a realization of an $N$-security portfolio
$P_i$ with the weights equal to the eigenvector components
$x_i^{(k)}, k=1,...,N$. For a non-degenerate matrix ${\bf C}$,
$P_i$ and $P_j$ are independent for each pair of their indices,
which allows one to choose such a portfolio, whose risk is
independent of others. According to the classical
theory~\cite{markowitz52}, the risk $R (P) = \sigma^2 (P) = {\rm
var} \{G_P(j)\}_{j=1}^T$ for the relevant group of securities can
be related to correlations (or covariances) between the time
series of individual security returns $g_s(j), j=1,...,T$.

Each eigenvector determined by Eq.~(\ref{diag}) (and thus
portfolio) can be associated with the corresponding time series of
the portfolio's returns by the expression
\begin{equation}
z_i(j,\Delta t) = \sum_{k=1}^N x_i^{(k)} g_k(j,\Delta t), \ \ i =
1,...,N; \ j=1,...,T, \label{eigensignals}
\end{equation}
which is analogous to Eq.~(\ref{portreturn}). These principal value
time series we shall call the eigensignals $Z_i$ (see
also~\cite{kwapien02,kwapien00} for some alternative
realizations). The risk associated with such eigensignals  is
related with the corresponding eigenvalues:
\begin{equation}
R (P_i) = \sigma^2 (Z_i) = {\bf x}_i^{\rm T} {\bf C} {\bf x}_i =
\lambda_i.
\label{eigenrisk}
\end{equation}
Thus, the eigenvalue size is a risk measure and, in consequence,
the larger $\lambda_i$, the larger variance of $Z_i$ and also the
larger risk of the corresponding portfolio $P_i$.

\section{Data specification}

This study of inter-stock correlations is based on high-frequency
data from the American stock market~\cite{data} in the period 1
Dec 1997 $-$ 31 Dec 1999. We chose a set of stocks of $N=100$
highly capitalized companies listed in NYSE or NASDAQ
(capitalization $> \$10^{10}$ in each case). These stocks are
sufficiently frequently traded (0.01-1 transactions/s) so that the
time scale of $\Delta t=5$ min allows to perform a statistically
significant analysis. For such a time scale the length of the time
series exceeds 40,000 data points. From the perspective of our
present purpose this time scale and the corresponding length of
the series turn out to constitute a reasonable compromise.
Thinking in terms of the Epps effect~\cite{epps79}, in the liquid
contemporary markets the time horizon of $\Delta t=5$ min is long
enough so that the cross-correlations get sufficiently expressed
beyond the noise level~\cite{kwapien04,borghesi07}. In case of the
data considered here the time horizon at which $\lambda_1$
saturates at its maximum corresponds to about 30 min, while for
$\Delta t=5$ min $\lambda_1$ assumes approx. 2/3 of its saturation
level. The length of the time series, on the other hand, allows one
to study the $T$ dependence of cross-correlations in a relatively
broad range of time intervals up to the maximum which corresponds
to $Q=406$.

\section{Data analysis}

One natural characteristics that may offer some introductory
insight when relating a given correlation matrix to the RMT is the
distribution of matrix elements. For our correlation matrices two
such distributions corresponding to the full $Q=406$ and to $Q=3$,
which in this latter case is obtained by properly windowing the
same time series and averaging over the windows, versus the best
Gaussian fits, are shown in Figure 1. Both these distributions are
shifted more towards positive values. In the case of $Q=406$ the
distribution is naturally much narrower than for $Q=3$ and shows
essentially no presence of negative matrix elements. This signals
that the real correlations are less contaminated by the
measurement noise for $Q=406$. Also the Gaussian fit in this case
is less satisfactory, especially in the region of larger positive
values of $C_{mn}$. Here, on the level of $1\%$ probability one
finds deviations of about $8\sigma$ (mean standard deviation)
while for $Q=3$ analogous deviations reach at most $3\sigma$.

\begin{figure}
\epsfxsize 10cm
\hspace{1.5cm}
\epsffile{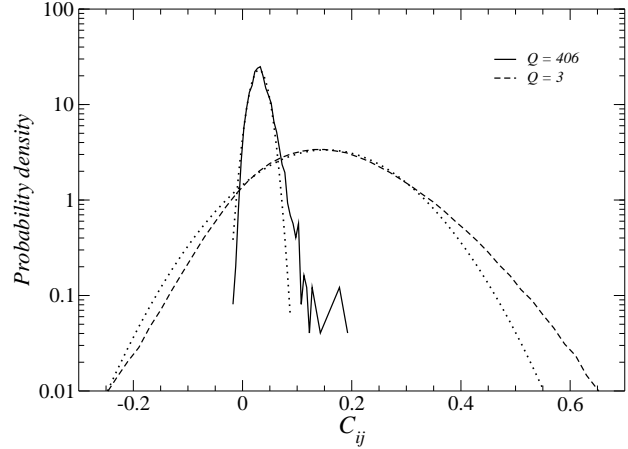}
\caption{Probability density distribution of entries of the
empirical correlation matrices ${\bf C}$ calculated for 100 highly
capitalized American companies over the period 1998-1999; the
solid line corresponds to $Q=406$ and the dashed line to $Q=3$.
The corresponding best Gaussian fits are indicated by the dotted
lines.}
\end{figure}

\section{Eigenvalue distribution}

A complementary and an even more informative characteristics of
the matrix is its eigenvalue distribution. Figures 2(a) and 2(b)
show all 100 eigenvalues distributed along the horizontal axis,
denoted by vertical lines, for the above presented cases of
$Q=406$ and $Q=3$, respectively. The largest eigenvalue
$\lambda_1$, assuming very similar values ($\sim 18-19$), repelled
from the rest of the spectrum, is seen in both cases and describes
the collective eigenstate which can be identified with the market.
The rest of the spectra develop however a significantly different
structure. For $Q=3$ this rest covers a much wider range of values
but at the same time its overlap with the corresponding random
Wishart matrices region (shaded vertical), whose bounds are
prescribed by the Eq.~(\ref{wishart}), is very substantial ($87\%$)
while for $Q=406$ it is rudimentary and looks pure coincidence. Of
course, concerning agreement of the empirical spectra with the RMT
this case of $Q=406$ is much more meaningful as compared to $Q=3$.
One more interesting, and probably related effect, is that for
$Q=406$ one sees (Fig.~2(a)) the second $\lambda_2$, and even the
third $\lambda_3$, eigenvalues that also are clearly separated
from the bulk. These eigenvalues can be related to some
branch-specific factors. No such factors can directly be seen for
$Q=3$ (Fig.~2(b)).

Due to the matrix trace conservation (here ${\rm Tr}{\bf C}=100$),
the existence of strong collective components can effectively
supress the noisy part of the ${\bf C}$ eigenspectrum, shifting
smaller eigenvalues towards zero and thus may distort their relation
to the RMT case. In order to correct for this effects, which more
is affecting the case of Figure 2(a), it is recommended to remove
the market factor $Z_1$ from the data~\cite{plerou02}. One way to
do this is by means of the least square fitting of this factor
represented by $z_1(j)$ to each of the original stock signals
$g_k(j)$:
\begin{equation} g_k(j)=\alpha_k +
\beta_k z_1(j) + \epsilon_k^{(1)}(j),
\end{equation}
where $\alpha_i,\beta_i$ are parameters, and then one can
construct a new correlation matrix ${\bf C}^{(1)}$ from the
residuals $\epsilon_k^{(1)}(j)$ (e.g.
ref.~\cite{laloux99,plerou02}). After this is performed
significantly more eigenvalues for $Q=406$ fall within the shaded
RMT region as Figure 2(c) illustrates. For $Q=3$ such a removal
does not affect so much the bulk of the original spectrum as
Figure 2(d) compared to 2(b) shows. Such a removal can be executed
once again and the $\lambda_2$ components can also be removed
leading to the eigenspectra presented in Figs 2(e) and 2(f). The
effect of this second removal is already much smaller but is more
noticeable in the former case.  In the corresponding Figure 2(e)
one still finds only ($\gamma=49$\%) eigenvalues overlapping with
the RMT interval $<\lambda_{\rm min},\lambda_{\rm max}>$. This is
almost a factor of two less than the case of $Q=3$ in Figure 2(f)
and only this latter result remains in agreement with results
presented earlier in~\cite{laloux99,plerou02} (based on the daily
data but with similar small values of $Q$) where a vast majority
of the eigenvalues was within the RMT bounds.

\begin{figure}
\epsfxsize 6cm
\hspace{-0.5cm}
\epsffile{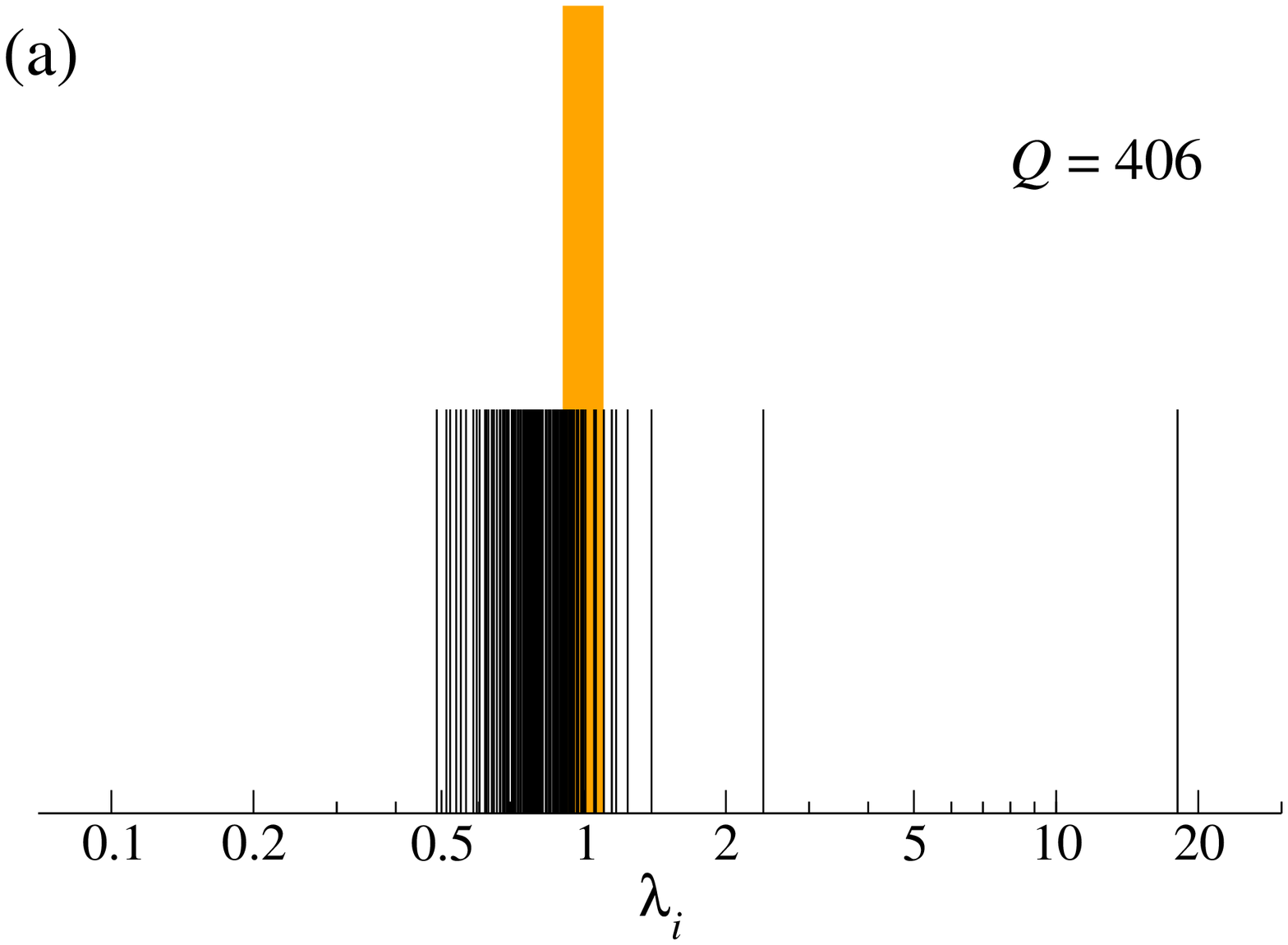}
\hspace{0.5cm}
\epsfxsize 6cm 
\epsffile{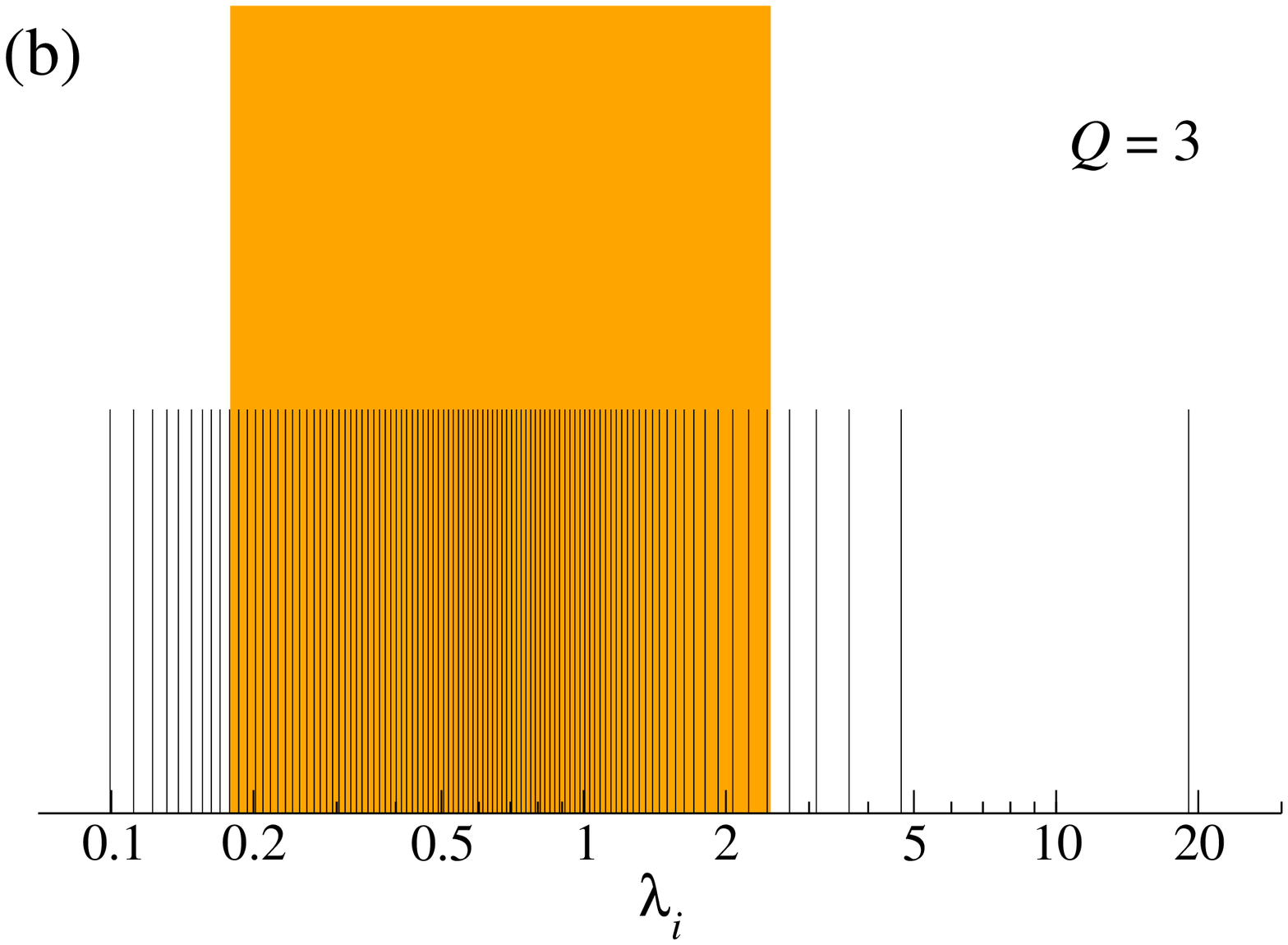}

\vspace{0.5cm}
\epsfxsize 6cm
\hspace{-0.5cm}
\epsffile{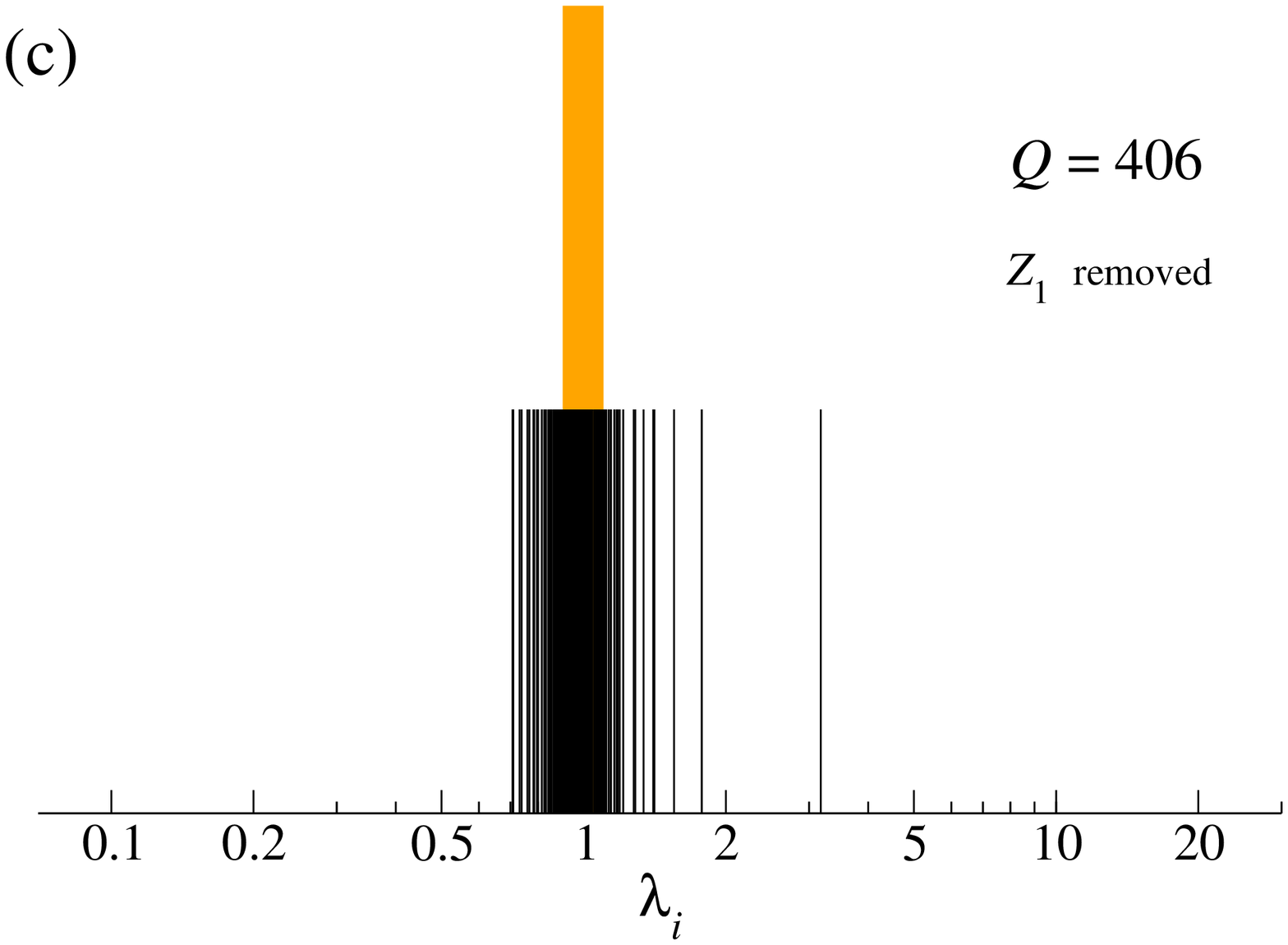}
\epsfxsize 6cm
\hspace{0.5cm}
\epsffile{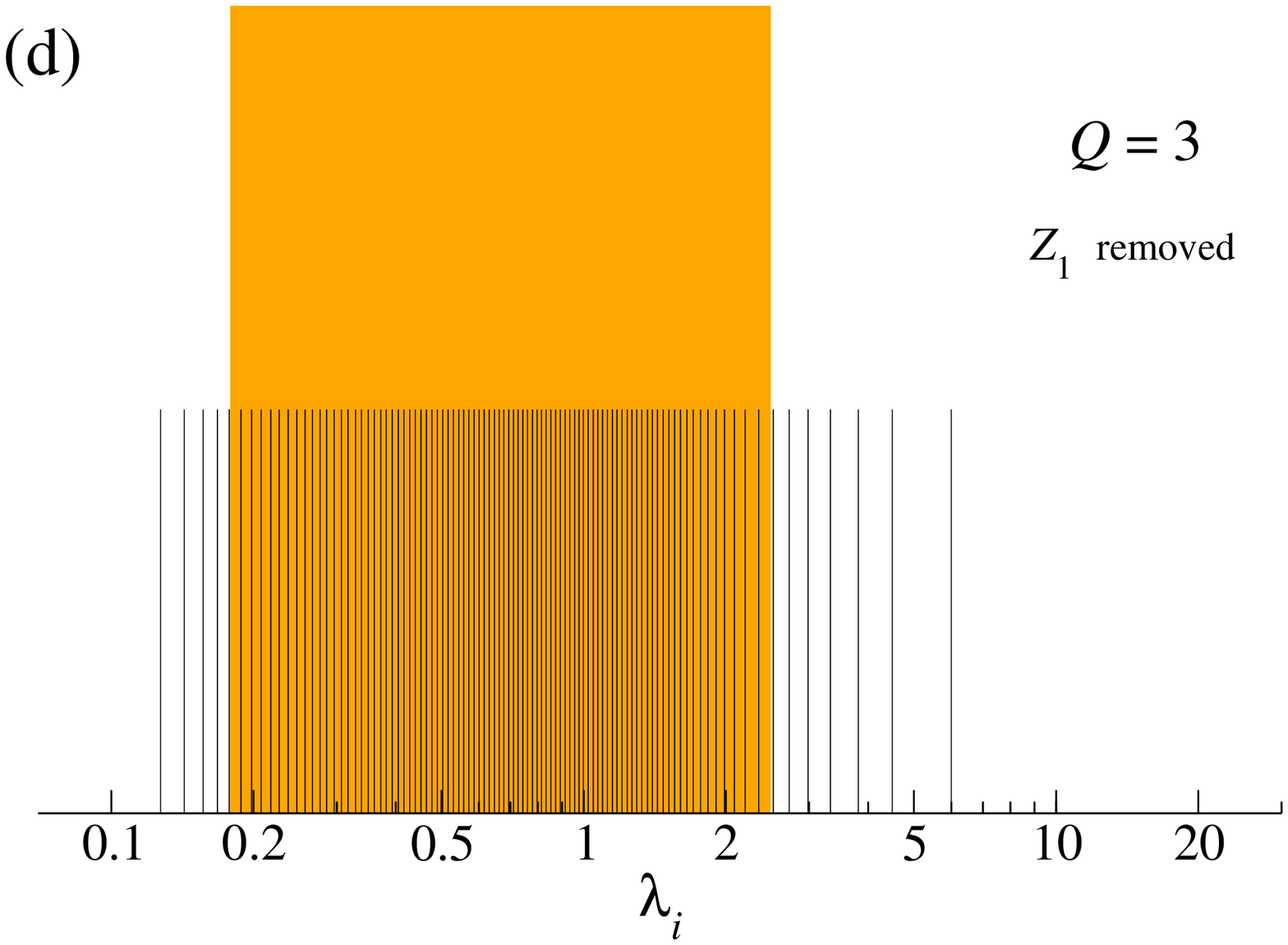}

\vspace{0.5cm}
\epsfxsize 6cm
\hspace{-0.5cm}
\epsffile{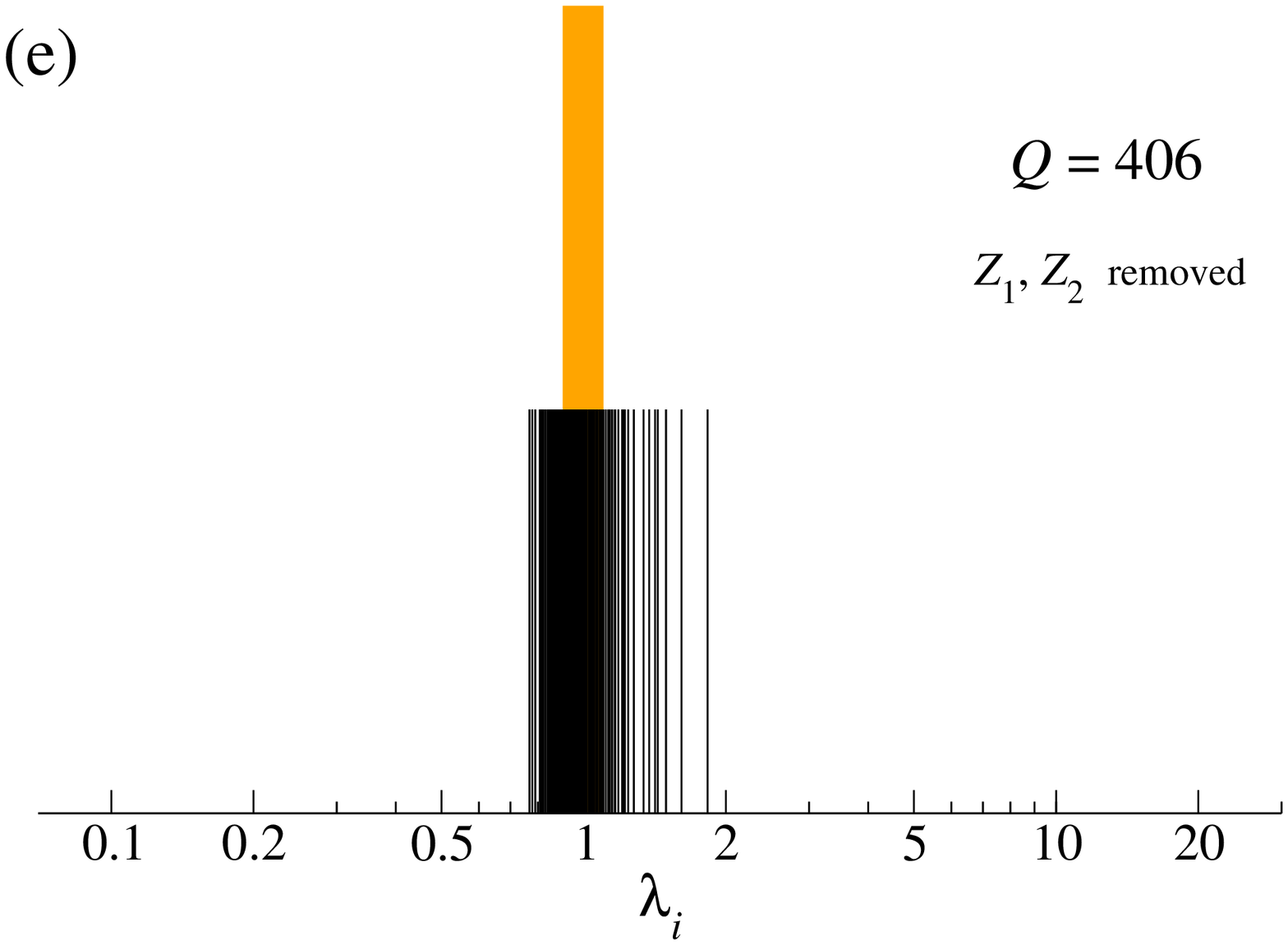}
\epsfxsize 6cm
\hspace{0.5cm}
\epsffile{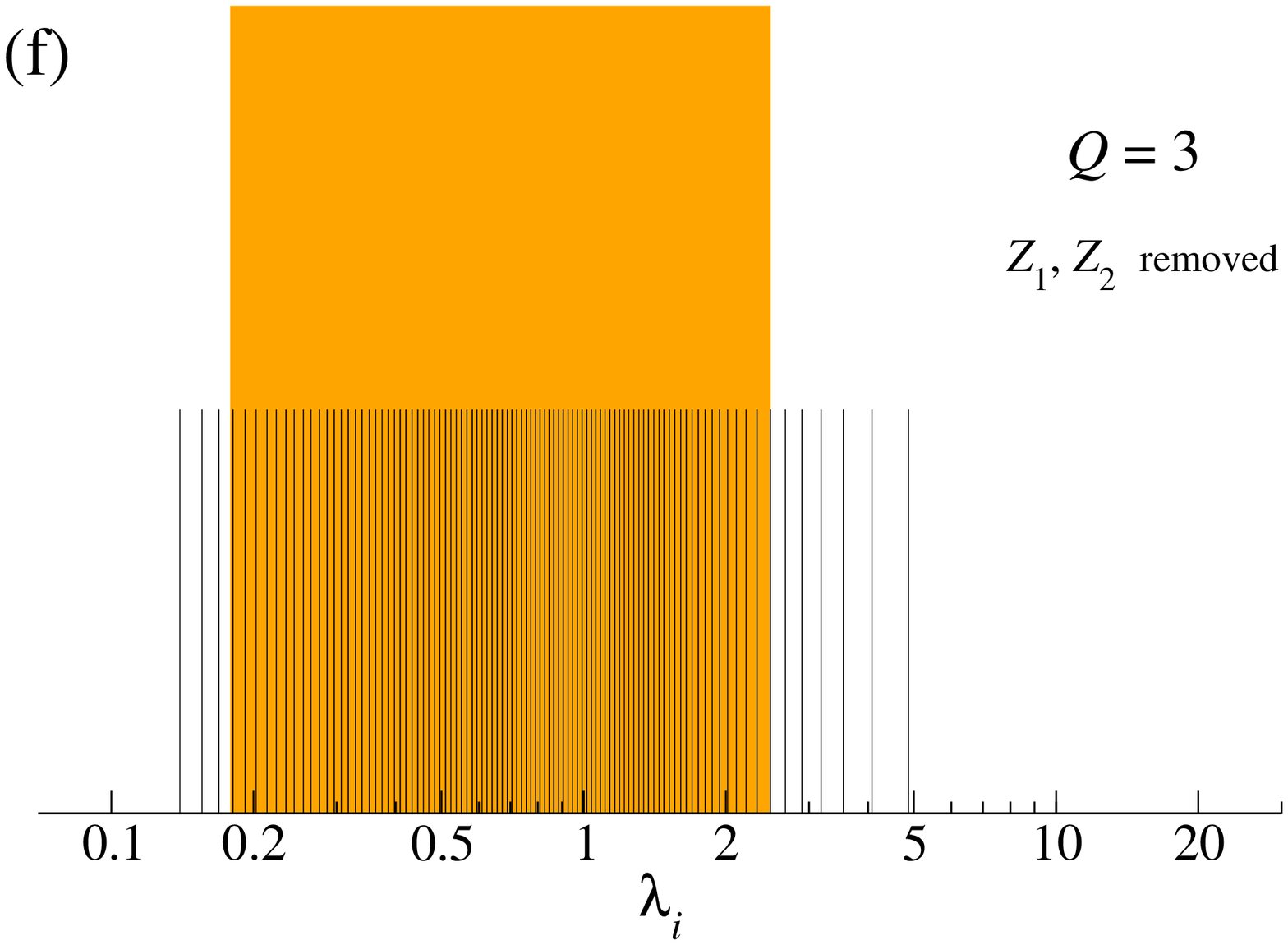}
\caption{Empirical eigenvalue spectrum of the correlation matrix
${\bf C}$ (vertical lines), calculated for 100 highly capitalized
American companies over the period 1998-1999 for $Q = 406$ (a) and
for $Q =3 $ (b); the eigenvalues of a random Wishart matrix with
the same $Q$ may lie only within the shaded vertical region.
Eigenvalue spectrum after effective rank reduction of ${\bf C}$,
i.e. after subtracting the contribution of the most collective
eigensignal $Z_1$ for $Q = 406$ (c) and for $Q = 3$ (d), and of the
two most collective ones $Z_1$ and $Z_2$ (c) for $Q = 406$ (e) and
$Q = 3$ (f).}
\end{figure}

\begin{figure}
\epsfxsize 6cm
\hspace{-0.5cm}
\epsffile{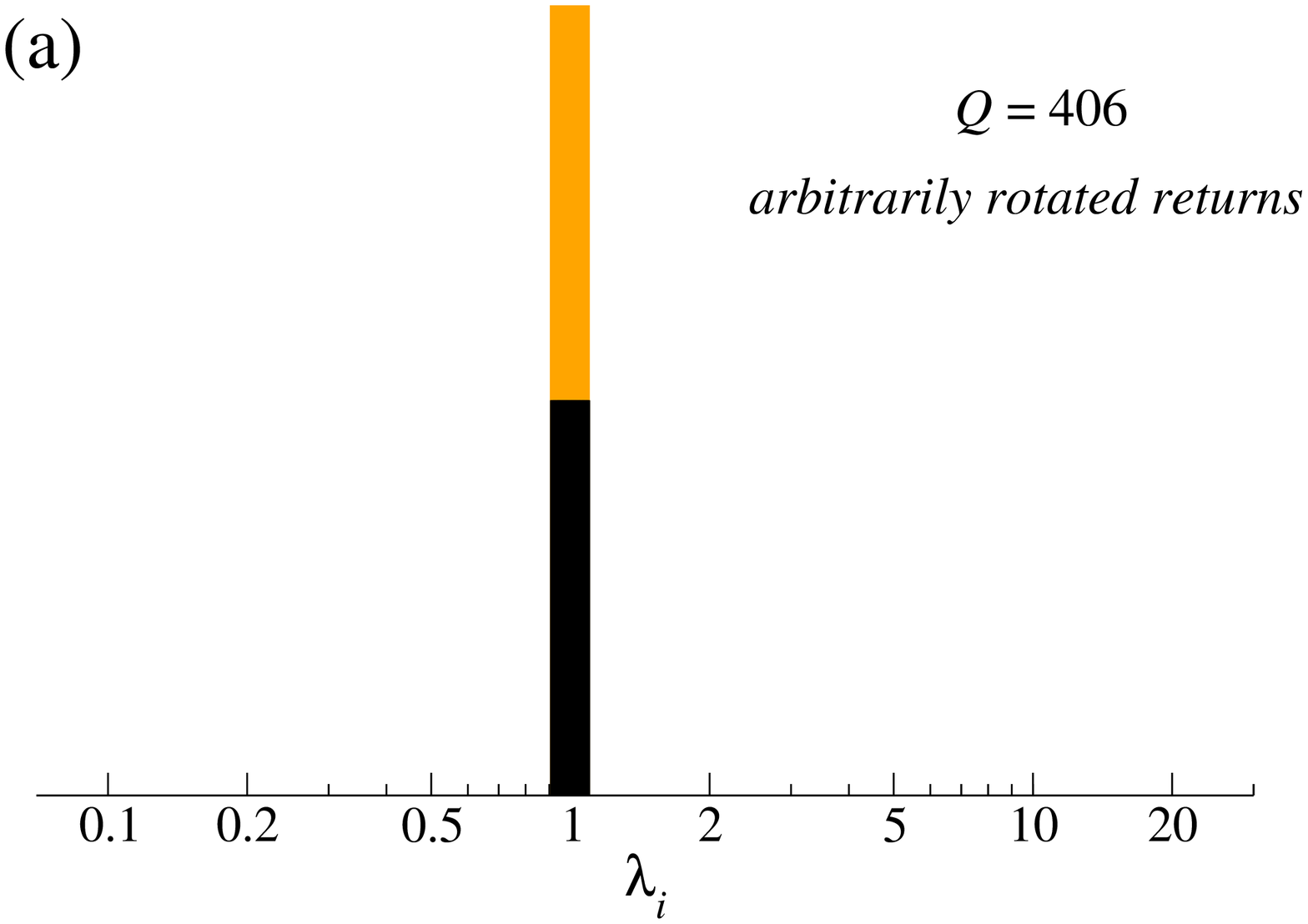}
\hspace{0.5cm}
\epsfxsize 6cm
\epsffile{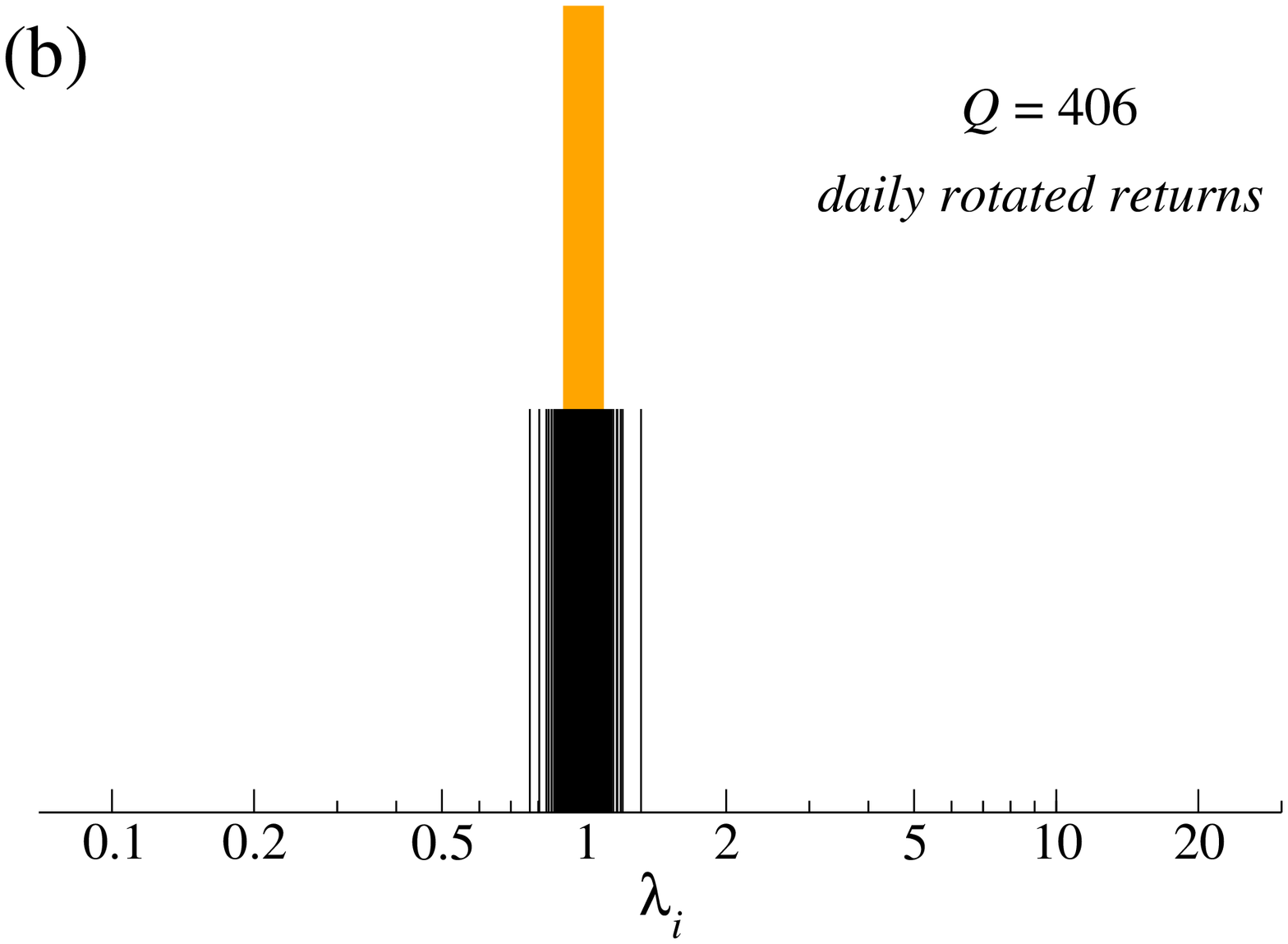}
\caption{Eigenvalue spectrum of the test correlation matrices for $Q =
406$ obtained after an unrestricted random shift (see text) of all the
original empirical time series against each other (a) and after
restricting this random shift to the multiples of one full trading day
(b). Shaded regions correspond to RMT predictions for the same value of
$Q$.}
\end{figure}

\section{Auxiliary tests}

There is potentially one effect that may partly be responsible for
such a sizeable disagreement between the $Q=406$ empirical and the
corresponding RMT results. There namely exists some time
correlations - especially the volatility correlations - in the
individual empirical time series that may effectively reduce the
number of independent events in each series. If this is the case
then the parameter $Q$ used in the reference RMT formula should
proportionally be smaller, the RMT bounds wider and thus an
agreement improved. In order to verify to what extent such an
effect may here be present we perform the following exercise.
Imagine all the time series are progressing along the independent
circles each, such that the end of the series is connected to its
starting point. The circles are then rotated against each other by
a random angle. This procedure preserves the internal correlations
within each series but destroys the cross-correlations. The
spectrum of eigenvalues of the so-randomized empirical correlation
matrix is shown in Figure 3(a). The perfect coincidence between
this empirical and the RMT result can be seen. This provides a
strong evidence that the corresponding disagreement in Figs.~2 is
entirely due to the real cross-correlations and is fully
informative.

As another related test the above circles are randomly rotated but
this time the rotation angles are restricted to the multiples of
one full trading day (daily rotated). Now, as is shown in Figure
3(b), the empirical spectrum broadens more than a factor of two
relative to the previous case and of course by the same factor
relative to the RMT bounds. This result may reflect the presence
of day-to-day repeatable intraday patterns of activity that affect
various different securities at similar instants of time during
the day.

\begin{figure}
\epsfxsize 6cm
\hspace{-0.5cm}
\epsffile{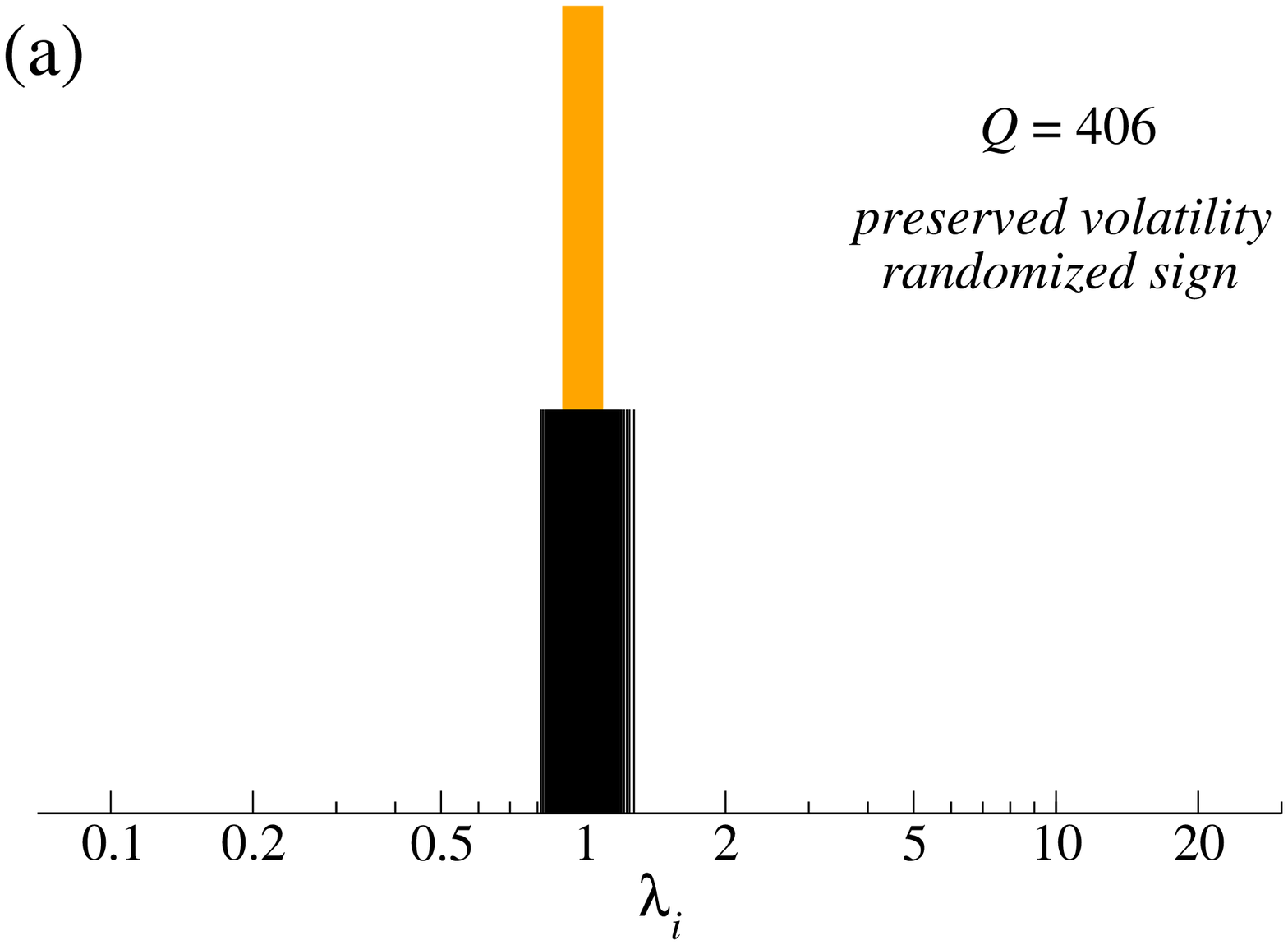}
\hspace{0.5cm}
\epsfxsize 6cm
\epsffile{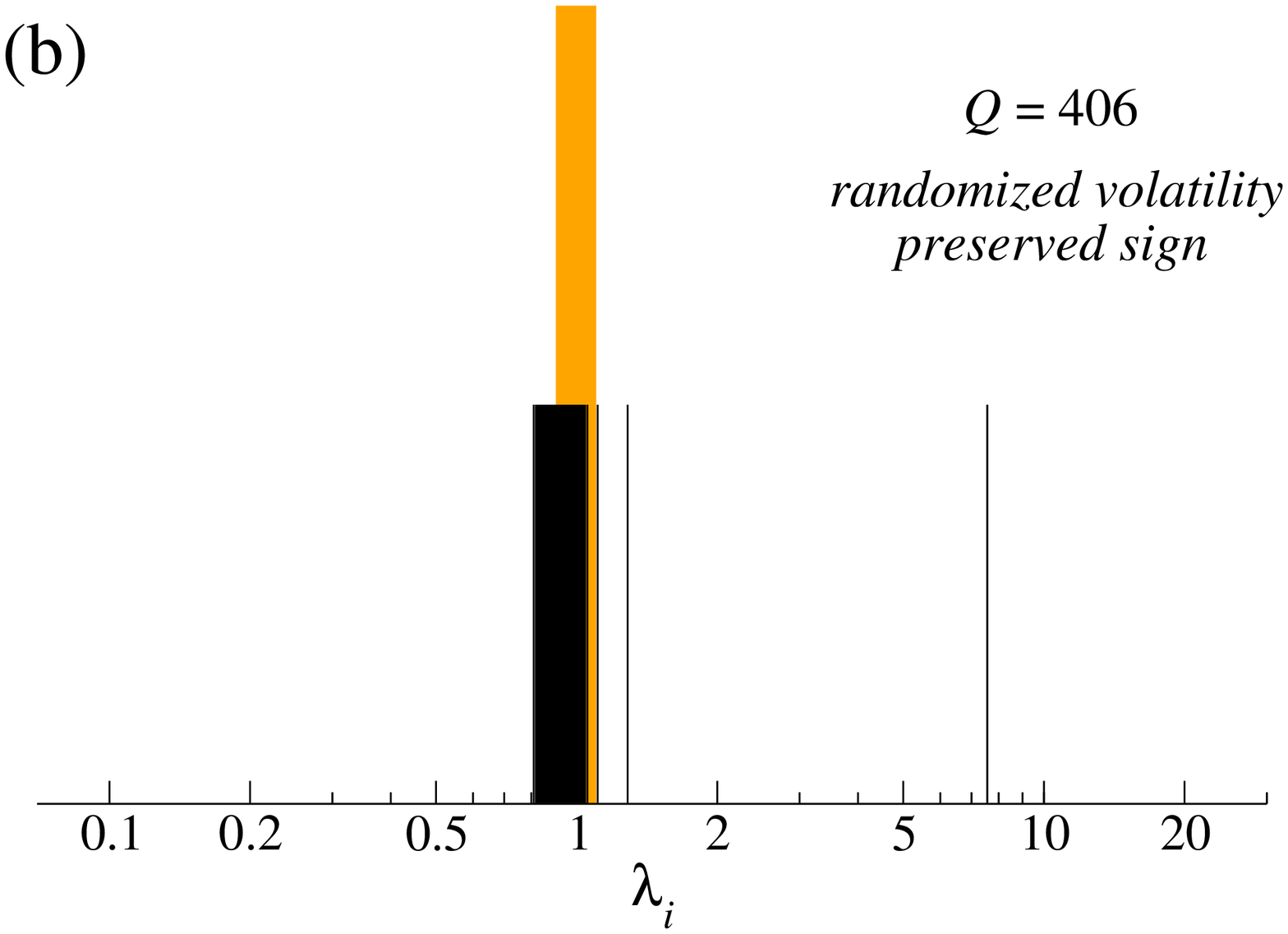}
\caption{Eigenvalue spectrum of the test
correlation matrices for $Q = 406$ obtained after the signs
(Eq.~(\ref{svreturn})) are randomly reshuffled, independently in
each empirical time series and the magnitudes are left unchanged
(a), and after the magntudes are randomly reshuffled and the signs
unchanged (b). Shaded regions correspond to RMT predictions for
the same value of $Q$.}
\end{figure}

As a further examination of the character of the stock market
cross-correlations two more types of artificial series based on
the original empirical data are created using the decomposition as
in Eq.~(\ref{svreturn}). Before the correlation matrix is calculated
either (a) the signs (${\rm sign}_s(j)$) are randomly reshuffled but
the return magnitudes $v_s(j)$ left at their original places or
vice versa (b), the signs are left original but $v_s(j)$
reshuffled for each series independently. The resulting spectra
are shown in Figure 4(a) and 4(b) correspondingly. From this
perspective the signs turn out responsible much more for the
cross-correlations than the corresponding magnitudes of the
returns. As is clearly seen, randomizing signs washes out the
cross-correlations almost completely (though deviations relative
to the RMT still remain) while randomizing $v_s(j)$ with the signs
unaltered largely preserves the original (Fig.~2(a)) structure of
the spectrum. To a good approximation the spectrum of Fig.~4(b)
looks compressed by a factor of about two relative to that in
Fig.~2(a).

\begin{figure}
\epsfxsize 6cm
\hspace{-0.5cm}
\epsffile{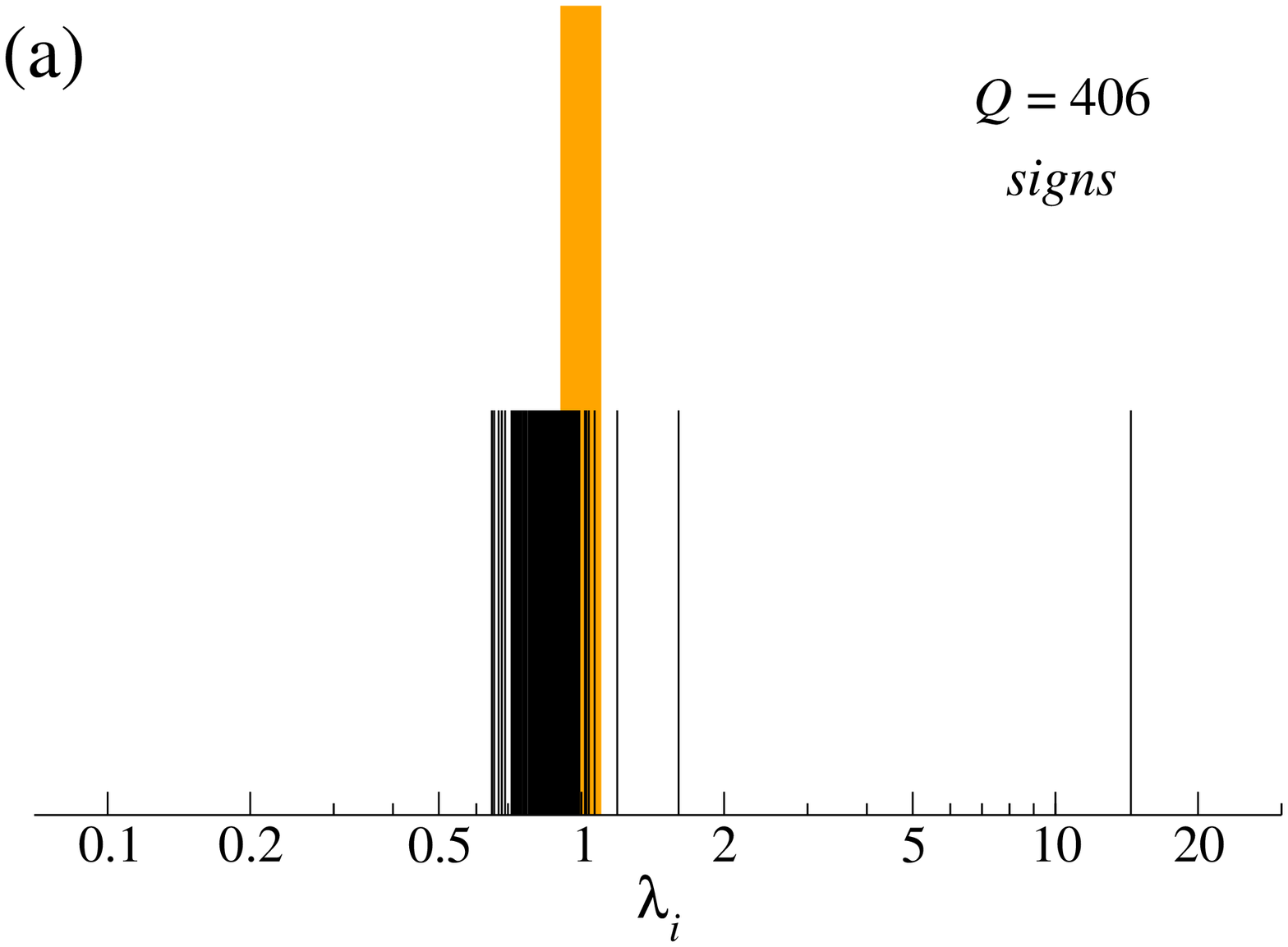}
\hspace{0.5cm}
\epsfxsize 6cm
\epsffile{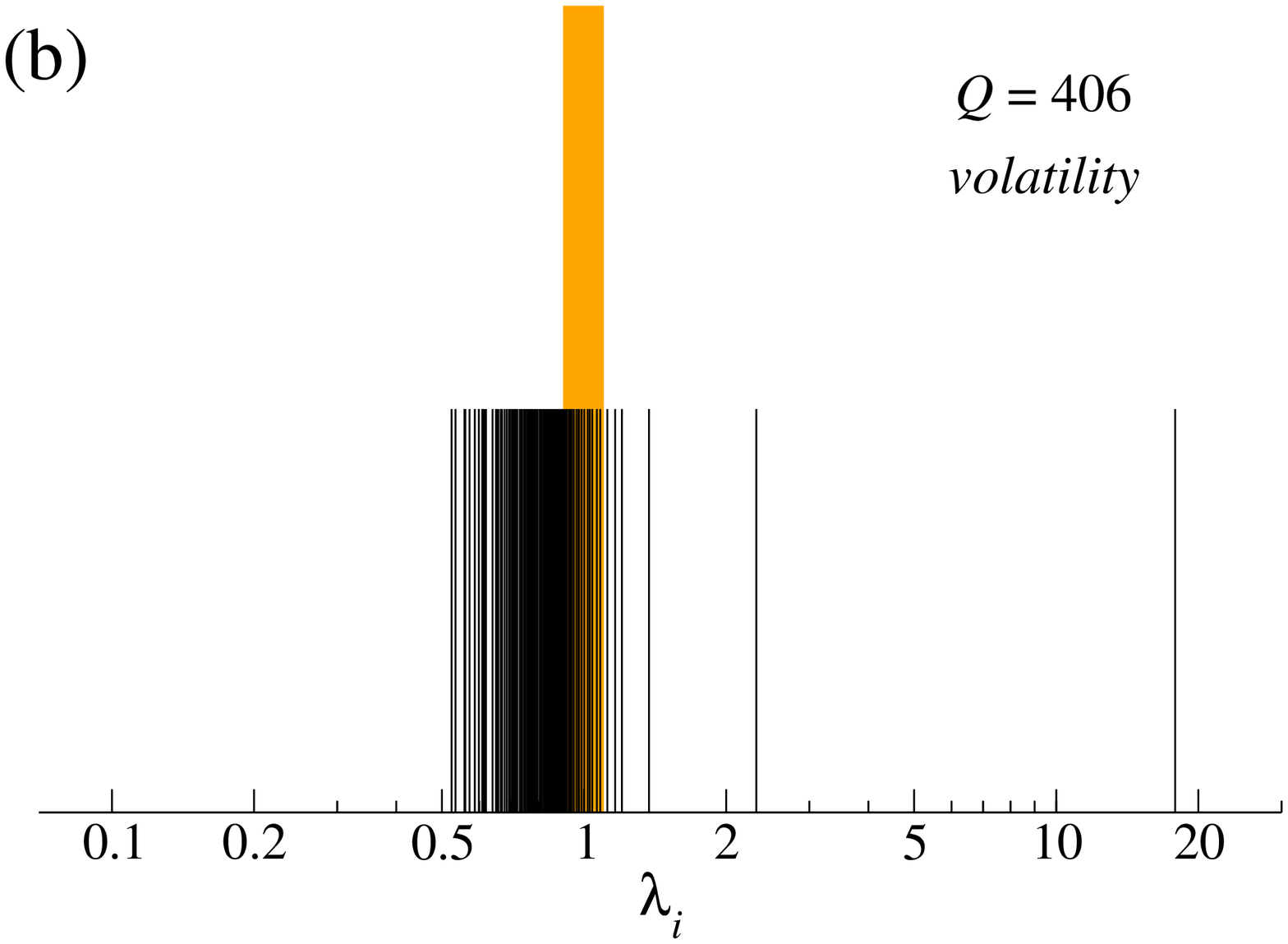}
\caption{Eigenvalue spectrum of the test correlation matrices for $Q=406$ 
obtained from the time series of signs (a) and from the time series of 
moduli of the empirical returns (b) as decomposed by Eq.~(\ref{svreturn}). 
Shaded regions correspond to RMT predictions for the same value of $Q$.}
\end{figure}

As a supplementary material to this kind of the test analysis in
Figure 5 we show the spectra (a) of the correlations matrices
calculated from the time series of ${\rm sign}_s(j)$ and (b) from the
time series of $v_s(j)$, independently. Consistently with the
observation made in Figure 4(b) the time series of the empirical
return's signs show very similar structure of cross-correlations
as the full original result (Figure 2(a)). In view of the result
presented in Figure 4(a) somewhat surprising may however be
considered the fact that the top eigenvalues appear (Figure 5(b))
even bit larger in the second case of $v_s(j)$ time series.
Relevant here is that these volatility related cross-correlations
manifest their presence only when the return's signs are entirely
discarded, i.e., their moduli are taken, which is a nonlinear
operation. The correlation matrix detects the linear (cross-)
correlations, but detecting linear correlations in volatility
means detecting the nonlinear correlations in returns. Thus the
above results taken together also point to the complex nonlinear
character of the financial cross-correlations.

\section{Eigensignal properties}

A deeper exploration of the relation between the characteristics
of the empirical financial cross-correlations and those of the
conventional RMT needs to involve also the eigensignals since they
directly reflect the dynamics of the corresponding portfolio.
Figure 6 presents the time series of the eigensignal returns
$z_1(j)$ calculated according to Eq.~(\ref{eigensignals}) for the
most collective eigenstate associated with $\lambda_1$ and for
another one associated with $\lambda_{52}$. Even though this latter
case corresponds to the middle of the empirical spectrum, strongly
overlapping with the RMT region, it appears difficult to detect
any significant differences, if one compares both series visually,
ignoring different scales in vertical axis. Both eigensignals are
nonstationary with likely extreme fluctuations and both of them
also exhibit volatility clustering. A compact form to quantify the
related effects is in terms of the multifractal spectra. It is a
well established fact that stock returns form signals which are
multifractal both on daily and on high-frequency time
scales~\cite{ivanova99,matteo04,oswiecimka05,kwapien05}.

\begin{figure}
\epsfxsize 10cm
\hspace{1.5cm}
\epsffile{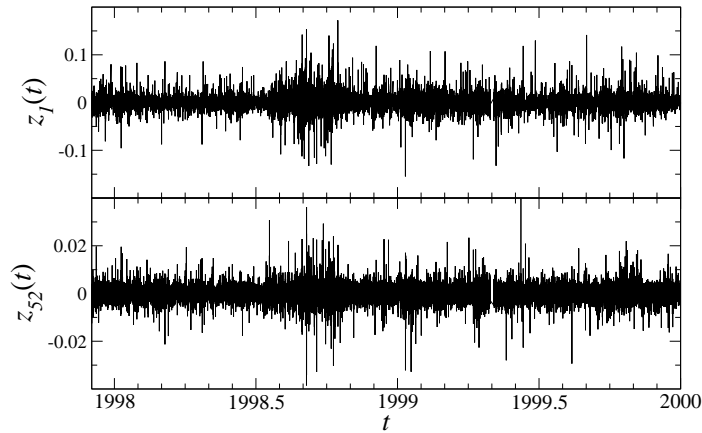}
\caption{Time series of the eigensignals for $\lambda_1$ (top) and
$\lambda_{52}$ (bottom). Note different scales in vertical axes of
both panels.}
\end{figure}

In order to evaluate the singularity spectra $f(\alpha)$ we use
the Multifractal Detrended Fluctuation Analysis
(MFDFA)~\cite{peng94} method which for the present purpose
appears~\cite{oswiecimka06} more stable than the Wavelet Transform
Modulus Maxima (WTMM) method~\cite{arneodo95}. Accordingly, we
start from our eigensignal $i$ represented by the time series
$z_i(j)$ of length $N_{es}$ and evaluate the signal profile
\begin{equation}
Y(j) = \sum_{k=1}^j{(z_i(k)-<z_i>)}, \ j = 1,...,N_{es},
\end{equation}
where $<...>$ denotes averaging over $z_i(k)$. In the next step
$Y$ is divided into $M_{es}$ segments of length $n$ ($n < N_{es}$)
starting from both the beginning and the end of the time series so
that eventually there are $2 M_{es}$ segments. In each segment
$\nu$ a local trend is removed by fitting an $l$-th order
polynomial $P_{\nu}^{(l)}$ to the data. Then, after calculating
the variance
\begin{equation}
F^2(\nu,n) = \frac{1}{n} \sum_{k=1}^n \{Y[(\nu-1) n+k] -
P_{\nu}^{(l)}(k)\}^2
\end{equation}
and averaging it over $\nu$'s, we get the $q$th order fluctuation
function
\begin{equation}
F_q(n) = \bigg\{ \frac{1}{2 M_{es}} \sum_{\nu=1}^{2 M_{es}}
[F^2(\nu,n)]^{q/2}\bigg\}^{1/q}, \ \ q \in \mathbf{R}
\end{equation}
for all values of $n$. For a signal of the fractal character
$F_q(n)$ obeys a power-law functional dependence on $n$:
\begin{equation}
F_q(n) \sim n^{h(q)}, \label{scaling}
\end{equation}
at least for some range of $n$. If this is the case the MF-DFA
procedure provides a family of generalized Hurst exponents $h(q)$,
which form a decreasing function of $q$ for a multifractal signal
or are independent of $q$ for a monofractal one. A compact form to
present the result graphically is to calculate the singularity
spectrum $f(\alpha)$ through the relations:
\begin{equation}
\alpha=h(q)+q h'(q) \hspace{1.0cm} f(\alpha)=q [\alpha-h(q)] + 1.
\label{singularity}
\end{equation}

\begin{figure}
\epsfxsize 10cm
\hspace{1.5cm}
\epsffile{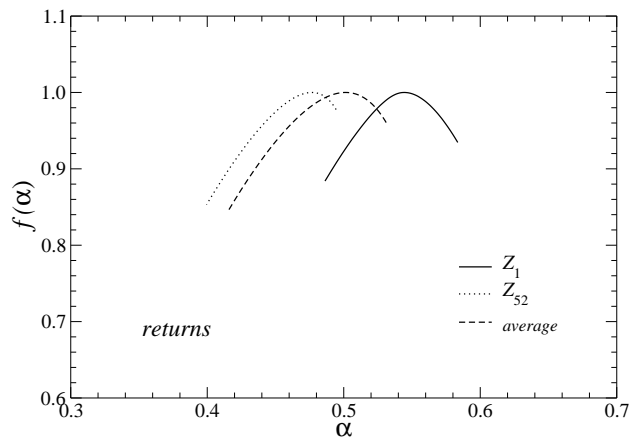}
\caption{(a) Singularity spectra $f(\alpha)$ for the eigensignal  
corresponding to the largest eigenvalue $\lambda_1$ (solid line), to the 
average over all other eigensignals $Z_i, \ i = 2,...,100$ (dashed line),
and to $Z_{52}$ (dotted line) of the empirical correlation matrix.}
\end{figure}

Some representative final results of such an analysis are shown in
Figure 7. Both eigensignals presented in Figure 6 ($Z_1$ and
$Z_{52}$) develop convincing multifractality with the spectrum
$f(\alpha)$ of about the same width even though the later one
represents the middle of the eigenvalue spectrum. The maxima of
these $f(\alpha)$ spectra are however located at different
positions, even relative to $\alpha = 0.5$, which may reflect
either persistency or antipersistency in the underlying time
series. As far as the width of $f(\alpha)$ is concerned they are
of comparable magnitude for all other eigensignals. As a global
documentation of this fact the average over all ($i= 2,...,100$)
the corresponding singularity spectra is also shown in this
Figure. This average displays maximum at $\alpha = 0.5$ exactly.

\section{Summary}

The results presented in this contribution provide further
evidence that the financial markets constitute a real complexity.
The stock market cross-correlations viewed through the
eigenspectrum of the correlation matrix show existence of the
market linear collective component represented by one pronounced
eigenvalue which is well separated from the bulk of eigenvalues.
This 'bulk' is however not of the Wishart random matrix ensemble
type which is especially clearly seen when the time series used to
construct the correlation matrix are sufficiently long. The fact
that the financial cross-correlations appear not to be simply
decomposable into 'market', 'sectors', and an uncorrelated Wishart
'bulk' has to do with their nonlinear character both in space and
in time. This profound nonlinearity manifests itself in the
multifractal nature of all the principal components (eigensignals)
which represent different portfolios and in the volatility
cross-correlations. This signals that information about the whole
market is encoded already in all its constituents. This does not
necessarily mean that the involved whole amount of information is
of practical interest or importance. In order however to
disentangle - in the spirit of the Random Matrix Theory - what is
more relevant from what is less, a more extended variant of random
matrix ensemble is called for. In view of the results presented
above, when postulating an appropriate RMT variant to be used as a
reference in the financial context one definitely needs to
redefine the notion of noise such that some of the correlations
are already built into.

\end{document}